\documentstyle[12pt]{article}
 \advance\oddsidemargin by -1in
 \advance\evensidemargin
 by -1in
 \marginparsep
 0.4cm
 \advance\topmargin by
 -0.0in
 \makeindex

 \newcommand\la{\langle}
 \newcommand\ra{\rangle}
 \newcommand\beq{\begin{equation}}
 \newcommand\noi{\noindent}
 \newcommand\eeq{\end{equation}}
 \newcommand\beqn{\begin{eqnarray}}
 \newcommand\eeqn{\end{eqnarray}}
 \newcommand{\doublespace} {
 \renewcommand{\baselinestretch} {1.6}
 \large\normalsize}

\begin{document}
\vspace*{1.5cm}
\hspace*{9cm}{\large MPIH - V32 - 1996}\\
\smallskip
\hspace*{9.65cm}{\large hep-ph/9607474}\\
 
\vspace*{4cm}
 
 \centerline{\Large \bf Baryon Asymmetry}
\medskip
 \centerline{\Large \bf of the Proton Sea at Low $x$}
 
\vspace{.5cm}
\begin{center}
{\large Boris~Kopeliovich\footnote{On leave of absence from Joint Institute
 for
 Nuclear
 Research, Laboratory of Nuclear Problems,
\newline Dubna, 141980 Moscow Region, Russia.  E-mail:
 bzk@dxnhd1.mpi-hd.mpg.de} and Bogdan Povh\footnote{E-mail:
 povh@dxnhd1.mpi-hd.mpg.de}}
 
\vspace{0.3cm}
 
 {\sl Max-Planck Institut f\"ur Kernphysik, Postfach
103980, 69029 Heidelberg, Germany}
 
\end{center}

\vspace{1cm}
 
\begin{abstract}
We predict a nonvanishing baryon asymmetry of the
 proton sea at low $x$. It
is expected to be about
 $7\%$ and nearly $x$-independent at $x < 0.5\times
10^{-3}$.  The asymmetry arises from the
 baryon-antibaryon component of the
Pomeron, rather
 than from the valence quarks of the proton, which
 are wide
believed carriers of baryon number.
 Experimental study of $x$-distribution of
the
 baryon asymmetry of the proton sea can be
 performed in $ep$ or $\gamma
p$ interactions at
 HERA, where $x\sim 10^{-5}$ are reachable, smaller
 than
at any of existing or planned proton
 colliders.
 
\end{abstract}

\doublespace
 
\newpage
 
\noindent
 {\large\bf 1. Introduction}
\medskip
 
 The carrier of the baryon identity, the baryon
 number (BN) is
defined rather loosely in most of
 the hadronic models, 
with a
 possible exception of the color string model.
In this paper we consider the partonic treatment of BN.
 
 One can prescribe $BN=\pm 1/3$ to each quark or
antiquark.  In this case BN asymmetry is a direct
 consequence of the
quark-antiquark asymmetry or
 vice versa.  Then the question arises, how can
BN
 of the proton find itself down at low $x$?  The
 simplest and wide spread
prejudice is, that BN is
 carried only by the valence quarks.  This means
that in order to find the proton BN at low $x$ one
 should slow down at least
one of the valence
 quarks to low $x$.  An example is shown
 schematically in
Fig.~1a, where a vertical axis is
 assumed to correspond to Bjorken $x$.
This
 mechanism provides BN with the same distribution
 $\propto 1/\sqrt{x}$
as for valence quarks.
 
\begin{figure}[tbh]
\includegraphics{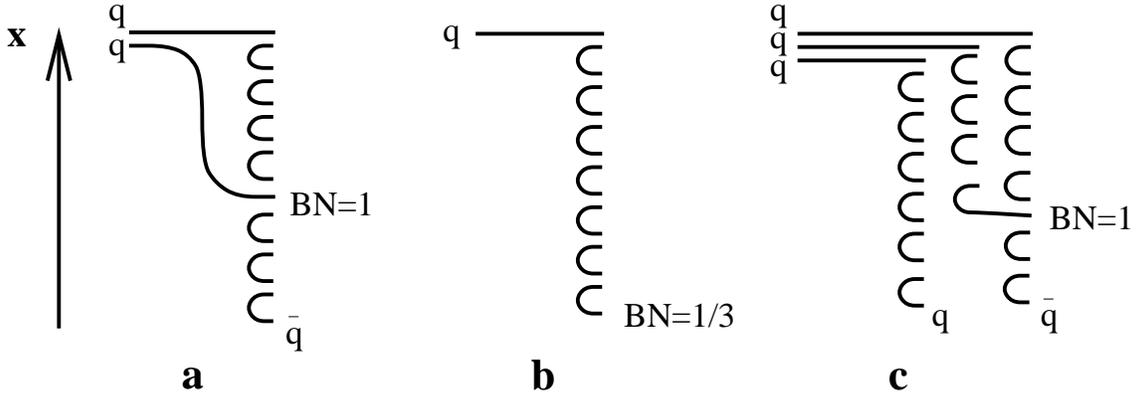}
\begin{center}
\vspace{5.5cm}
\parbox{13cm}
 {\caption[Delta]
 {\setlength{\baselineskip}{0.55cm}
 \it
 Different mechanisms of BN flow down to low
 $x$.  {\bf a:} The valence
 quark
 itself is slowed down.
 {\bf b:} The valence quark surrounded by a
 sea-parton
 cloud.
 BN flows down to low $x$ through the $q\bar q$ chain.
 {\bf c:}
 Three valence quarks in a decuplet-color state
 develop three
 $q\bar q$
 chains, which carry the BN.
 Conventionally we assume a
 vertical
 $x$-axis.}
\label{fig1}}
\end{center}
\end{figure}

 There is, however, another way to transport the
 $BN=1/3$ of the valence
quark down to low $x$
 through a chain of sea quarks - antiquarks as is
illustrated in Fig.~1b.  Since the chain is
 BN-symmetric, except the very
last quark at low
 $x$, the valence BN finds itself at low $x$.  It
 is
natural that in this case the BN has the same
 distribution $\sim 1/x$ as the
sea quarks.
 
 An important difference from the previous example
 in Fig.~1a
is that there is no correlation between
 the flavour of the low-$x$ quark
carrying the BN
 and the flavour of the valence quark initiating
 the chain.
In this sense one may say that not only the
 valence quark itself, but the $q -
\bar q$ chain
carries the BN as well.
 
 However, the two other valence quarks
in the
 proton (in any baryon) are mostly in a
 color-antitriplet state and
develop their own
 $q\bar q$ chain, but of an opposite alignment:
 anti-BN
$-1/3$ is transported down to low $x$,
 where it compensates exactly the BN of
the sea of
 the first valence quark.  This fact makes it
 difficult to
realize the mechanism of BN flow
 shown in Fig.~1b.
 
 There is, however, a
probability to find two
 valence quarks in the proton in a sextet-color
state. It may happen in the higher Fock
 components of the proton, for
instance in
 $|uudgg\rangle$.  The two additional gluons can be
 in different
color-states, a singlet, two octets and
 an antidecuplet.  In the latter case
any of the
 two valence quarks are in a color-sextet state,
 i.e.  each of
the valence quarks develops its own
 sea $q\bar q$ chain and transports its BN
to low
 $x$ as illustrated in Fig.~1c, where the gluons 
responsible for color conservation are
not shown. We estimate the
 probability of such a
configuration below.
 
\medskip

One can interpret these results in terms of string 
model as well. In this approach 
baryons are assumed to have a
Y-star configuration. This suggests to relate BN
 with the
 string junction,
the point where the three strings
 join \cite {rv}. In this case BN is carried
by
 gluonic
 field rather than by the valence quarks at the
 endpoints of
the strings.  Indeed, if the baryon
 is excited, each of the strings may break
due to
 $q\bar q$ pair production.  In this case the
 valence quarks split
away as mesons, but
 eventually a baryon is produced around the same
 string
junction.
 
 Such a star-structure of baryon naturally arises
 as a string
analogue to the locally
 gauge-invariant operator with $BN=1$ \cite{rv}
 
\beqn
 |3q^v\rangle
 &=&J^{i_1i_2i_3}(X)
 G_{i_1}^{j_1}[P(X,X_1)]
G_{i_2}^{j_2}[P(X,X_2)]
 G_{i_3}^{j_3}[P(X,X_3)]\times\nonumber\\
 &
&q^v_{j_1}(X_1)
 q^v_{j_2}(X_2)
 q^v_{j_3}(X_3)\ ,
\label{1}
\eeqn
 
 \noi
 where
 
 \beq
 G_{i}^{j}[P(X,X')] =
 \left[T\
\exp\left\{ig\int_{P(X,X')}
 A_{\mu}(X)dX^{\mu}\right\}\right]_i^j
\label{2}
\eeq
 
 \noi
 the integration goes along path $P(X,X')$
 between points
$X$ and $X'$.  Tensor
 $J^{i_1,i_2,i_3}(X)$ should be associated with
string
 junction $J$ having coordinate $X$ as is
 illustrated in Fig.~2a.

 As soon
 as the string junction shares the proton
 momentum it is
reasonable to provide
 it with a
 partonic interpretation.  Let us consider
the
 Fock-state
 decomposition of the light-cone wave
 function of the
proton.
 
 \beq
 |p\rangle = |3q^v\rangle + |3q^vq^s\bar q^s\rangle +
|3q^v2q^s2\bar
 q^s\rangle + ...
\label{3}
\eeq
 
\begin{figure}[tbh]
\includegraphics{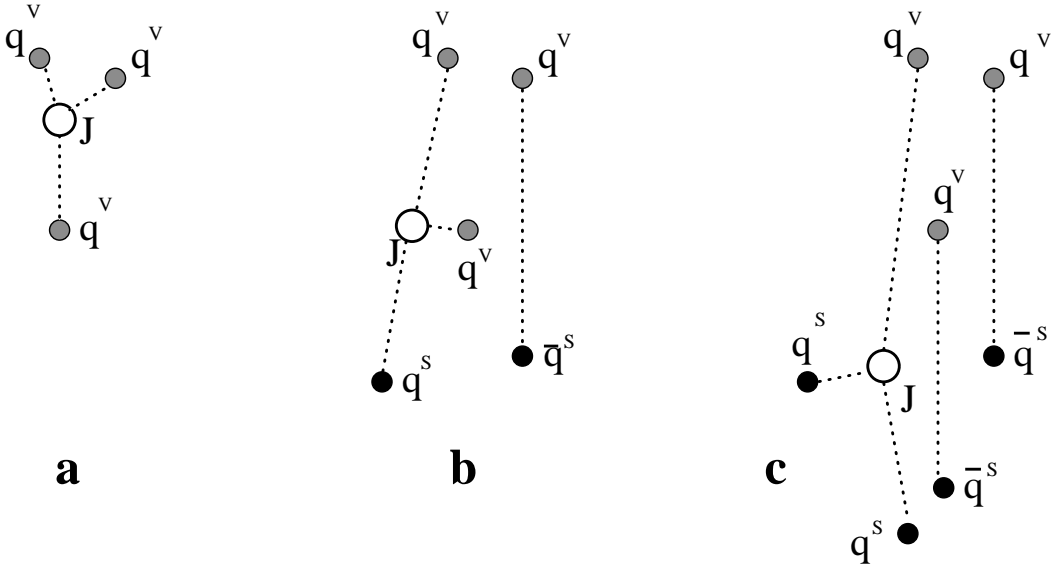}
\begin{center}
\vspace{7.5cm}
\parbox{13cm}
 {\caption[Delta]
 {\setlength{\baselineskip}{0.55cm}
 \it The
 cartoon shows the string configurations
 which correspond to different terms
 in Fock state
 decomposition (\ref{3}).  Grey and black circles
 show the
 valence and sea quarks respectively.  The
 open circles show the position of
 the string
 junction.  The dotted lines correspond to triplet
 color
 strings. Conventionally we assume a
 vertical $x$-axis.}
\label{fig2}}
\end{center}
\end{figure}

 \noi
 The {\sl first term} of (\ref{3}) corresponds to
 expression
 (\ref{1}) and is presented in Fig.~2a.
 In order to move the string junction
 down to lower
 $x$ one should slow down at least two of the three
 valence
 quarks.  We assume dominance of a minimum
 energy configuration of the
 strings, what makes
 the string junction to follow the diquark rather
 than
 a single quark.  Thus, the probability to
 find the string junction at low
 $x$ in this Fock
 state is 
 
 \beq
 B_1(x)
 \propto
 \frac{1}{x^{\alpha^0(M^J_4)}}\ ,
\label{1a}
\eeq
 
 \noi
 where $\alpha^0(M^J_4)$ is the intercept of the
 Regge
trajectory corresponding to
 diquark-antidiquark mesons ($M^J_4$ according
to
 the classification and notations in \cite{rv})
 shown schematically in
Fig.~3a.
 
\begin{figure}[tbh]
\includegraphics{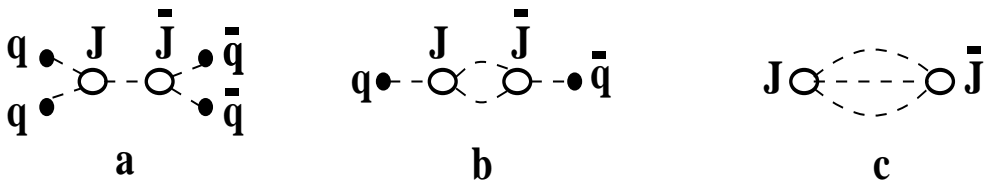}
\begin{center}
\vspace{2.5cm}
\parbox{13cm}
 {\caption[Delta]
 {\it Mesonic states (notations from
 \cite{rv})
 lying on the Regge trajectories
 giving rise to BN exchange.
 {\bf a:} The
 diquark-antidiquark meson $M^J_4$.
 {\bf b:} The $qJ - \bar
 q\bar J$ meson
 $M^J_2$.
 {\bf c:} The $J - \bar J$ glueball $M^J_0$.}}
\label{fig3}
\end{center}
\end{figure}

 It is related to known
 parameters of other
 Reggeons,
 
 \beq
 \alpha^0(M^J_4) = 2\alpha^0_N -
 \alpha^0_{\rho}
 \approx -1.2\ ,
\label{1b}
\eeq
 
 \noi
 where $\alpha_N \approx -0.4$ and $\alpha_{\rho}
 \approx
0.5$ are the intercepts of the nucleon and
 $\rho$-meson trajectories.
Although relation
 (\ref{1b}) was first derived in \cite{eh} using an
oversimplified multiperipheral model, it also
 follows from the factorization
relations of
 \cite{kaidalov} for planar graphs (see also review
\cite{capella}).
 
 The {\sl second } Fock component in (\ref{3}) can
 be
split into two colorless clusters, a $3q$-star
 and $q\bar q$ color dipole.
The corresponding
 wave function reads
 
 \beqn
 |3q^vq^s\bar
 q^s\rangle
&=&
 q^v_{j_1}(X_1)
 G_{i_1}^{j_1}[P(X,X_1)]
 J^{i_1i_2i_3}(X)
G_{i_2}^{j_2}[P(X,X_2)]\
 q^v_{j_2}(X_2)\
 G_{i_3}^{j_3}[P(X,X_3)]\
q^s_{j_3}(X_4)\nonumber\\
 &
 &\bar
q^{sm}(X_5)G_{m}^{l}[P(X_3,X_5)]q^v_{l}(X_3)\ ,
\label{4}
\eeqn
 
 \noi
 where we have chosen the case when the dipole
 contains
the
 valence quark and the sea antiquark.
 Of course such a specific color
distribution may
 be suppressed by a color factor, which affects
 only the
probability, but not the $x$-dependence.
 The string configuration
corresponding to the wave function
 (\ref{4}) is shown in Fig.~2b. The
minimum
 energy string
 star in the $|2q^vq^s\rangle$ state
 corresponds
to
 the string junction having nearly
 the same $x$ as the slowest of the
valence
 quark.
 This is shown in Fig.~2b, where we assume
 conventionally
a
 vertical
 $x$-axes.
 
 Thus we come to the conclusion that the 5-quark
Fock component
 of the proton light-cone wave
 function provides a low-$x$
distribution of
 the
 string junction which follows that for the valence
quarks, i.e.  $\sim
 1/\sqrt{x}$.
 
 On the other hand, the low-$x$
behaviour of the string
 junction
 accompanied by a valence quark
corresponds to the intercept of the Regge
 trajectory for the $qJ-\bar J\bar
q$ states
 ($M^J_2$ in notations of
 \cite{rv}) shown in Fig.~3b.
 
 \beq
B_2(x)
 \propto
 \frac{1}{x^{\alpha^0(M^J_2)}}\ .
\label{5a}
\eeq
 
 \noi
 We conclude from the above consideration that
 
 \beq
\alpha^0(M^J_2) = {1\over 2}\ .
\label{5}
\eeq
 
 This value of the intercept first claimed in
 \cite{kz3,kz4}, is higher than one 
 suggested in \cite{eh,rv} on
 the basis of an oversimplified
 multiperipheral bootstrap model. 
There are, however, a
 few experimental confirmations of the
 value given in (\ref{5}).
 
\begin{itemize}

\item
 It is demonstrated in \cite{kz4} that (\ref{5}) is
 in perfect
 agreement with the energy dependence of
 the data \cite{isr1,isr2,isr-rev}
 on
 BN
 number production
 in the central rapidity region in $pp$
 collisions.
 The
 absolute value of the cross section evaluated
 in
 perturbative QCD in
 \cite{kz4} agrees with the
 data as well. Below we
 present a new
 calculation,
 which does not rely upon pQCD.
 
\item
 Evaluation of the $\bar pp$
 annihilation cross section \cite{kz3}
 assuming dominance of the $M^J_2$ Reggeon exchange
 is in a good agreement
 with available (up to $12\
 GeV$) data. The value of the intercept
 eq.~(\ref{5})
 naturally explains the observed energy
 dependence
 $\sigma^{p\bar p}_{ann} \propto 1/\sqrt{s}$.
 
\item
 The dynamics of the $M^J_2$ Reggeon exchange
 was applied
 recently
 \cite{ck} to BN stopping in
 high-energy heavy ion collisions.  It
 was
 found to
 be a dominant mechanism for net BN production at
 mid
 rapidities
 at $200\ GeV$ and nicely explains the data
 from the NA35
 experiment
 \cite{na35} on S-S
 collisions.  The predicted baryon stopping
 in
 Pb-Pb
 collisions \cite{ck} was confirmed recently
 by data
 \cite{klmm}
 at $158\
 GeV$.
 
\end{itemize}

 The {\sl third} term in eq.  (\ref{3}) is
 illustrated in Fig.~2c, where
 we
 again include a
 valence quark and a sea antiquark in each
 color
 dipole.
 The wave function of this Fock component
 reads
 
 \beqn
 |3q^v2q^s2\bar
 q^s\rangle &=&
 q^v_{j_1}(X_1) G_{i_1}^{j_1}[P(X,X_1)]
 J^{i_1i_2i_3}(X)
 G_{i_2}^{j_2}[P(X,X_2)]\
 q^s_{j_2}(X_4)\
 G_{i_3}^{j_3}[P(X,X_3)]\
 q^s_{j_3}(X_6)\nonumber\\
 &\times &\bar
 q^{sm}(X_5)G_{m}^{l}[P(X_2,X_5)]q^v_{l}(X_2) \bar
 q^{sm}(X_7)G_{m}^{l}[P(X_3,X_7)]q^v_{l}(X_3)
\label{6}
\eeqn
 
 We again assume that the sea quarks have smaller
 $x$-values than
the valence ones, as is
 indicated in Fig.~2c.  The minimum energy
 of
the string configuration is reached when the
 string junction has nearly the
same $x$ as that of the two sea quarks.  Consequently, the
 string
junction,
 i.e. BN, has in the third term
 of Fock decomposition (\ref{3})
the
 same
 $x$-distribution $\sim 1/x$ as the sea quarks.
 
 This
important
 conclusion
 means that the Regge trajectory
 corresponding to the
$J-\bar J$
 mesons
 (Fig.~3c)
 $M^J_0$ (notation of \cite{rv})
 and
providing
 $x$-distribution of
 BN
 
 \beq
 B_3(x) \propto
\frac{1}{x^{\alpha^0(M^J_0)}}\ ,
\label{7}
\eeq
 
 \noi
 has intercept
 
 \beq
 \alpha^0(M^J_0) = 1\ .
\label{8}
\eeq
 
 \noi
Thus, we arrived at the same conclusion as we drew
from the consideration of the $q\bar q$ chains shown in Fig.~1.
 This result (\ref{8}) also follows from energy
independence
 of the $\bar pp$ annihilation cross section claimed in
\cite{gn},
 where it was assumed that
 BN annihilation results from overlap
of
 the string junction and antijunction in impact parameter plane.
 It was
guessed in analogy with inelastic reactions
 initiated by crossing of the
strings,
 that the string rearrangement in annihilation
 is also energy
independent. The authors of \cite{gn} estimated
also the asymptotic
annihilation cross
 section at $\sigma^{\bar pp}_{ann} \approx 1-2\ mb$.
 
Perturbative QCD calculations
 \cite{k,kz1,kz2,kz6} of the annihilation
 via
two-gluon exchange proved energy-independence
 of the cross section.
Moreover, even the absolute value of
 asymptotic annihilation cross
section
 was predicted in a parameter-free way
 to be the same, $1-2\ mb$, as
in \cite{gn}.
 
 The annihilation cross section is
 measured only up to
$12\
 GeV$, and it is not very
 likely to get data at much higher
energies. Nevertheless, a solid
 confirmation of the above
 predictions was
found in \cite{kz3} from an
 analysis of particle multiplicity distribution
in
 $\bar pp$ and $pp$
 interactions at high energies.
 The string
junction
 exchange, or its
 perturbative
 analogue the color-decuplet
gluonic
 exchange
 \cite{k}-\cite{kz2} lead to a three-sheet topology
 of
final state
 (see Fig.~1c). This
 implies a high multiplicity
 of produced
particles,
 about $3/2$ of the mean
 multiplicity. Such a signature allows
to
 single
 out
 a pure string junction
 exchange. Analysis
 \cite{kz2}
of
 available
 data on multiplicity distribution
 confirms the energy
independence of the
 cross
 section
 with value $\sigma^{p\bar p}_{ann} =
1.5 \pm 0.1\ mb$ in a
 perfect
 agreement with
 the theoretical
predictions
 \cite{gn,kz1,kz2}
 
 Note that these theoretical and
experimental
 results
 are in variance with
 the expectation of
\cite{eh,rv}
 that $\alpha^0(M^J_0) = 1/2$. 

\medskip

 Summarizing, we
 expect
the BN density distribution
 in the
 proton to have the form
 
 \beq
B_p(x) - \bar B_p(x) =
 \sum_{k=0}^2\ C_{2k}\
 x^{-\alpha(M^J_{2k})}
\label{9}
\eeq
 
 \noi
We subtract the anti-BN density in order to remove
the trivial baryon symmetric part of the sea.
 The sum rule representing BN conservation demands
 
 \beq
\int\limits_{x_{min}}^1 dx\left[B_p(x)
- \bar B_p(x)\right] = 1\ ,
\label{9a}
\eeq
 
 \noi
 where $x_{min} = Q^2/2m_N\nu$ or $m_N/\nu$ for virtual or
real
 photons, respectively.
 
In the present paper we suggest an
experimental
 study of the
 BN distribution at low $x$ in $ep$ or
 $\gamma
p$ interactions at HERA.  In
 order to
 study the momentum distribution of
the produced
 net BN one should
 measure the difference between
 the baryon ($B$) and antibaryon 
 ($\bar B$) production rates. Such a
 net BN distribution
reflexes the
 $x$-distribution of the string
 junction in
 the projectile
proton, since the
 final
 state baryon is produced with nearly
 the same
$x$.
 Note that $x$ is
 defined as a ratio of the final
 baryon to
 the
initial proton light-cone
 momenta,
 $x=p^+_B/p^+_{p}$, but not through the
virtuality
 and the energy
 of the photon. We predict
$d\sigma(\gamma^*p\to
 BX)/dx\propto 1/\sqrt{x}$
 at
 $x>5\times 10^{-4}$
and asymptotic behaviour,
 $d\sigma(\gamma^*p\to
 BX)/dx\propto 1/x$ at
smaller $x$.  HERA seems to be
 the best machine
 for
 such studies, since it 
provides the smallest
 values of
 $x$ compared to any of
 planned proton
colliders, RHIC, LHC or even SSC would
 have reached.

We discuss the ways to probe the baryon asymmetry of the
sea in the next section. We demonstrate that 
usual probe of the quark/antiquark asymmetry 
cannot be used at low $x$ and suggest to measure the 
baryon asymmetry of produced particles.

In section 3 we discuss and provide a numerical evaluation
of the gluonic contribution to the baryon asymmetry,
which dominates at very low $x$. 

The valence quark contribution to the baryon asymmetry,
which is important down to quite low $x$, is evaluated 
in section 4.

In section 5 we estimate unitarity corrections to BN distribution,
which may be important at high energies. We found a $20\%$
correction for the energy of HERA.

\bigskip

 \noi
 {\large\bf 2. How to probe the baryon
 asymmetry?}
\medskip
 
 BN, like gluons, cannot be directly
 probed by a virtual photon, and one
should look for other probes.

A baryon asymmetry of the sea obviously manifests itself
in the quark/antiquark asymmetry. The latter was
suggested in \cite{brodsky} to be
measured in deep-inelastic neutrino
interactions, which are different for
quarks and antiquarks. This method, however,
cannot be used to measure the low-$x$ baryon asymmetry
under discussion.
Indeed, the color string configuration shown in Fig.~2c
is quark/antiquark symmetric at low $x$, in spite of
the presence of the string junction (the $q/\bar q$ asymmetry
appears as a result of hadronization of the strings). 
A high-energy neutrino, which develops a $q_1\bar q_2$
($u\bar d,\ c\bar s$...) fluctuation through the $W$-boson,
interacts with this string configuration
with the same cross section as an antineutrino, as far as
the valence quark of the proton are not involved in
the interaction. Thus, it is insensitive to a baryon asymmetry. 
The same can be demonstrated in
the quark-chain representation illustrated in Fig.~1c. As for
$u\bar d$ and $\bar ud$ fluctuations,
the $\nu/\bar\nu$ symmetry is obvious
(provided that isospin symmetry
of the sea at low $x$ is true).
There is no symmetry, however, for strange quarks.
Since strangeness is conserved, its distribution  
has a maximum at
the rapidity of the string junction, and a negative
minimum in the vicinity in the $x$-scale.
Such an oscillation causes a complete cancellation
of the strangeness at a fixed value of $x$, probed
by the neutrino, when one averages over the $x$-value
of the baryon. This cancellation does not take place
in the case of quark/antiquark asymmetry generated
by $K\Lambda$ Fock component of the proton considered 
in \cite{brodsky}. 

 Searching for
 another signature of BN one can use
 the
shortness
 of rapidity-correlations
 between the
 primordial BN and the
produced baryon,
 typical for
 all known
 models of hadronization.  Therefore, we
assume that the
 $x$-distribution of the
 produced
 baryon is close to the
primordial BN
 distribution. Of course
 the baryon-antibaryon
 pairs
spontaneously produced
 from vacuum also
 contribute, but this background
can
 be eliminated
 by
 subtraction of baryon
 and antibaryon production
rates. We
define the baryon/antibaryon
production asymmetry as
 
 \beq
 A_B(x) =
 \frac{\Delta_B(x)}
 {\Sigma_B(x)}\ .
\label{9b}
\eeq
 
 \noi
 Here we denote $\Delta_B(x) = N_B(x)-N_{\bar
 B}(x)$ and
$\Sigma_B(x) = [N_B(x)+N_{\bar B}(x)]/2$,
 where $N_B(x) =
[xd\sigma(B)/dx]/\sigma_{in}$ is
 the ratio of the inclusive (anti)baryon
production
 to the total inelastic cross sections.
 
\vspace{0.7cm}
 
 \noi
 {\large\bf 3. Gluonic contribution to
the BN density at low $x$}
\vspace{0.3cm}
 
 Thus, the observable reflecting the BN
 distribution in
the proton is the baryon asymmetry
 (\ref{9b}) of produced particles. It
should be pointed out
 that partonic interpretation is not Lorentz
 invariant
and may look
 quite differently depending
 on the reference frame.
 For
example, one cannot say
 to which one of the two colliding
 hadrons the sea
parton belongs, as
 the answer depends on the
 reference frame.
 Even a
valence quark of one colliding hadron may look in
 the rest frame of
 this
hadron as a sea quark of
 another one.
 
 The
 same
 is true for the
partonic interpretation of
 the BN distribution in the
 proton.
 The
Lorentz-invariant observable, the baryon asymmetry
 of the
 produced
particles
 can be calculated, of course, in any
 reference frame. It is, however, most convenient to do 
the estimation in the proton rest frame, since one
 can use
available
 information on the BN annihilation
 cross section at high
energies, which we mentioned in
 the introduction. In this
 reference frame
the produced BN is supposed to preexist as
 a fluctuation of
 the photon,
accompanied by
 an anti-BN, due to BN conservation.
The latter has to annihilate with the BN
 of
 the
target in order to produce the observed
 baryon asymmetry. Two
 examples
are
 sketched in Fig.~4.
 
\begin{figure}[tbh]
\includegraphics{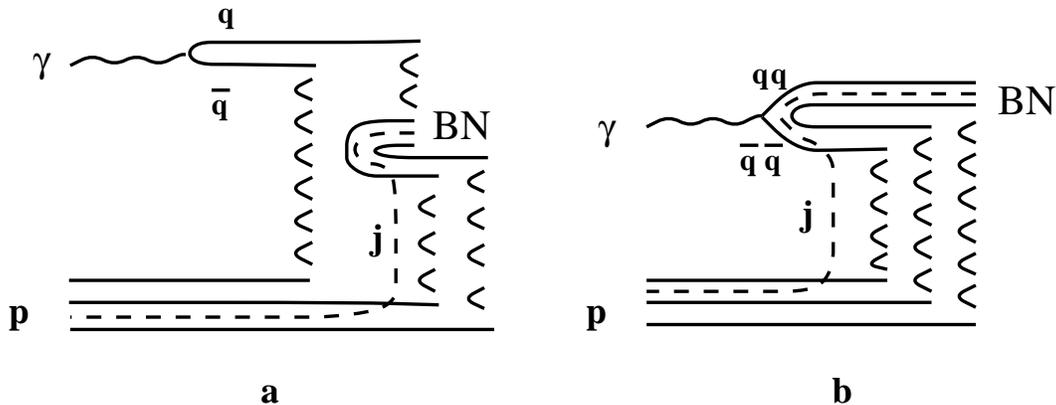}
\begin{center}
\vspace{6cm}
\parbox{13cm}
 {\caption[Delta]
 {\it Gluonic mechanism of the proton BN flow
 to
 the central rapidity region ({\bf a}) and to the
 photon fragmentation
 region ({\bf b}). The dashed
 lines show the trajectory of the string
 junction}}
\label{fig4}
\end{center}
\end{figure}

The amount of the sea $B\bar B$ pairs stored in the
photon fluctuation cancels in the 
relative asymmetry (\ref{9b}), and we arrive at a very simple
expression for the baryon asymmetry
 
 \beq
 A_B(x) =
 \frac{\sigma^{B\bar
 B}_{ann}(s=m_N^2/x)}
{\sigma^{hp}_{in}}\ ,
\label{10}
\eeq
 
 \noi
 which is very important for further
 applications.
Here $\sigma^{hp}_{in}$ is the
 inelastic cross section for the dominant
hadronic
 fluctuation of the photon at $s=m_N^2/x$.  It is
 the $\rho$-meson
in the case of a real photon, so
 we will use $\sigma^{hp}_{in}\approx 20\
mb$.  We
 do not expect any strong $Q^2$-dependence of the
 baryon asymmetry,
despite the fact that the
 photoabsorption cross section for highly virtual
photons decreases as $1/Q^2$.  This may be
 interpreted as a suppression $\sim
1/Q^2$ of
 interaction of small-size, $\propto 1/Q^2$,
 fluctuations of the
photon.  At the same time the
 baryon-antibaryon component of these
fluctuations
 has to have a small transverse separation as well.
 Thus the
annihilation cross section acquires the
 same suppression factor $1/Q^2$.
 
In order to proceed further with the evaluation of
 the baryon asymmetry
(\ref{10}) one needs to know
 the baryon-antibaryon annihilation cross
section
 $\sigma^{B\bar B}_{ann}$ at high energies.  As
 mentioned in the
introduction, the asymptotic behaviour  of
the annihilation cross section was studied in
nonperturbative \cite{gn} and perturbative
 \cite{kz1,kz6} QCD approaches, and
also analysing
 data on multiplicity distribution in $pp$ and
 $p\bar p$
interactions \cite{kz2,kz6}. Using so
 different ideas all these approaches
arrive at the
 same conclusion: the annihilation cross section
 at high
energies is about $1-2\ mb$ and nearly
 energy-independent.
 
 Using
(\ref{10}) we find $x$-independent baryon
 asymmetry $A_B^0 \approx 7\%$. This
asymmetry is
 due to flow of the BN of the initial proton
traced by the gluons.
 
 We can also
estimate the absolute value of the
 yield of the net BN $\Delta_B(x)$ provided
that
 the total yield of baryons is known.
 Baryon-antibaryon pair production
from vacuum is
 known to be substantially suppressed compared to
 mesons
\cite{cnn,book,diquarks}, $\Sigma_B(x) =
 \epsilon\ N_{\pi}(x)$, where $\epsilon
\approx
 0.08$ and $N_{\pi}(x) = [x
 d\sigma(\pi)/dx]/\sigma_{in}
\approx 3-5$
 \cite{tables}, dependent on energy.  Then
 eq.~(\ref{10}) leads
to the estimate
 
 \beq
 \Delta_B(x)
 \approx
 \epsilon\
\frac{\sigma_{ann}}{\sigma^{h p}_{in}}
 \ N_{\pi}(x)
 \approx
 0.02\ .
\label{16}
\eeq
 
 Thus, only about $2\%$ of the total
 photoabsorption cross section
goes for production
 of net BN per unit of rapidity.
 
 The realistic
rapidity distribution of particle
 production is nearly constant only in the
central region, but decreases towards
 the rapidity of the projectile.
 Thus,
eq.~(\ref{16}) may
 overestimate baryon production in the
 photon
fragmentation region. On the other hand, there is
 a specific
 channel of BN
production
 by means of a spontaneous
 dissociation of the photon into the
diquark-antidiquark pair.
 The anti-string-junction
 may subsequently
annihilate with
 the string junction of
 the proton, as sketched in
Fig.~4b.
 
 To evaluate the cross section of
 the net BN production
 in the
photon fragmentation region we take into
 account
 the suppression by factor
of $\sim 0.08$
 for the diquark compared to a
 quark
 pair
 production, and
the smallness of annihilation
 compared to
 the
 total inelastic cross
section,
 $\sigma_{ann}/ \sigma_{in}^{hp}
 \approx
 0.07$.
 Therefore, in
$0.6\%$ of all DIS events
 the net BN is produced
 by the photon
dissociation
 mechanism. The
 corresponding
 contribution to the inclusive
cross
 section
 $d\sigma(B-\bar B)/dy \propto
 exp(y-y_{\gamma})$ peaks at
the photon
 rapidity
 $y_{\gamma}$.
 
 Thus, the
 gluonic mechanism
 of
BN transfer
 predicts a plateau for net BN distribution
 at mid
rapidities
 and a peak
 in the photon fragmentation
 region.
 
\vspace{0.7cm}
 
 \noi
 {\large\bf 4. Valence quark contribution to
 the
BN
 distribution}
\vspace{0.3cm}
 
 We expect a substantial growth of the baryon
 asymmetry
towards the proton fragmentation region
 due to the quark mechanism of BN
transfer
 \cite{kz4}. On the other hand, it may extend down
to quite low $x$.
Perturbative calculation of the BN
 flow over large
rapidity intervals performed in
 \cite{kz4} is in a good agreement with the
measurement of the baryon asymmetry in central
 rapidity region measured in
$pp$ interaction at
 ISR \cite{isr1}.  The $x$-dependence of this
mechanism is controlled according to (\ref{5a}) -- 
(\ref{5}) by
 the leading Reggeon
intercept, $\alpha_R(0) =
 1/2$, so one can write 
\cite{kz4}
 
 \beq
 \Delta_B^{(q)}
=
 \delta_q\sqrt{x}\ .
\label{17}
\eeq
 
 \noi
 The factor $\delta_q$ can be either borrowed from
 the
calculations \cite{kz4}, or fixed by
 comparison with available data
\cite{isr1,isr2,isr-rev}, which gives $\delta_q
 \approx 0.6$.
 Assuming
that
 $\sigma^{pp}_{in} \approx 1.5\ \sigma^{\pi
 p}_{in}$ we get
 $\delta_q
\approx 1$
 
 One can also evaluate the contribution
 of the valence
quarks
 to the baryon asymmetry using same
 equation (\ref{10}) except the
annihilation
 cross section is to be evaluated
 within the same
 quark
mechanism of BN transfer as it was done in
 \cite{kz3}, or one can use
directly the data on
 annihilation cross section at
 preasymptotic
energies, $\sigma^{p\bar p}_{ann} \approx 70\ mb\
 \sqrt{s_0/s}$. Thus,
(\ref{10}) gives
 similar value $\Delta_q
 \approx 1.1$.
 
 The calculated
baryon asymmetry
 
 \beq
 A_B(\eta) = \frac{\Delta_B^{(q)}(\eta)}
{\Delta_B^{(q)}(\eta) +
 \epsilon N_{\pi}(\eta)}\ ,
\label{18}
\eeq
 
 \noi
 is plotted in Fig.~5 as
 function of the baryon rapidity
$\eta=\eta_p-ln(1/x)$ in the laboratory
 frame. From this figure we expect
that
 the gluonic mechanism dominates at $\eta < -1$ at
 HERA.
 
\begin{figure}[tbh]
\includegraphics{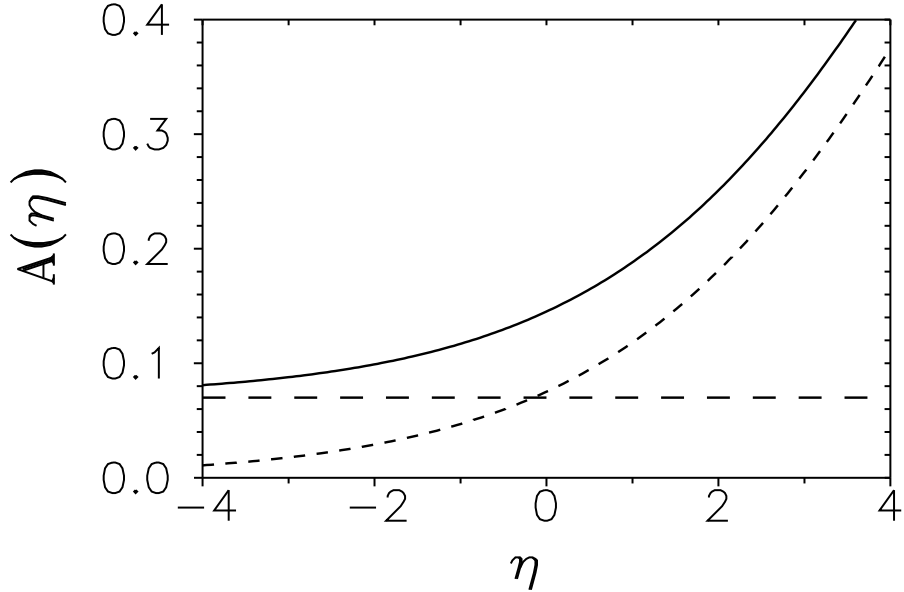}
\begin{center}
\vspace{7.5cm}
\parbox{13cm}
 {\caption[Delta]
 {\it Baryon asymmetry in $\gamma p$
 interaction at
 HERA versus rapidity in the laboratory frame.  The
 dashed
 line corresponds to the
 rapidity-independent gluonic mechanism of BN
 transfer.  The dotted curve represents the quark
 mechanism of BN transfer,
 calculated with
 (\ref{17})-(\ref{18}). The solid curve is the sum
 of the
 two contributions.}
\label{fig5}}
\end{center}
\end{figure}

\vspace{0.7cm}
 
\noindent
 {\large\bf 5. The BN conservation sum rule. Unitarity
 corrections.}
\vspace{0.3cm}
 
 In spite of the smallness of the baryon asymmetry
 it
comes to a contradiction with the BN
 conservation sum rule eq.~(\ref{9a}) at
very low
 $x$ if $A_B(x)=A_B^0$ is a constant.  Indeed,
 relation (\ref{9a})
for BN distribution can be
 rewritten for the baryon asymmetry using
(\ref{9b}) as
 
 \beq
 \int\limits_{x_{min}}^1
 \frac{dx}{x}\
 A_B(x)\
\Sigma_B(x) = 1\ ,
\label{19}
\eeq
 
 \noi
 but the left-hand side of this relation grows with
 $1/x$
since $\Sigma_B(x) \approx \epsilon
 N_{\pi}(x)$ and
 
 \beq
\int\limits_{x_{min}}^1
 \frac{dx}{x}\ A_B(x)\
 N_{\pi}(x) >
 A_B^0\ \la
N_{\pi}\ra\ ,
\label{20}
\eeq
 
 \noi
 where $\la N_{\pi}\ra = \int dx/x\ N_{\pi}(x)$ is
 the mean
multiplicity of produced
 pions, which is known to
 grow as $1/x^{\alpha_P
-
 1}$
 if the Pomeron intercept $\alpha_P > 1$,
 or as $\ln(1/x)$ for
$\alpha_P
 =
 1$. In any case the sum rule
 eq.~(\ref{19}) is violated.
 
The source
of the puzzle can be understood as
 follows.  Following section 3 we 
treat the baryon
 asymmetry as a result of creation of a sea
baryon-antibaryon
 fluctuation in the projectile
 photon and annihilation of
the antibaryon with
 the
 target proton. This mechanism leads to the basic
relation
 Eq.~(\ref{10}).  Although the probability
 of a $B\bar B$ fluctuation is
strongly suppressed
 by the smallness of the
 factor $\epsilon$, the number
of such pairs grow with $1/x$,
 and the
 amount of $B\bar B$ pairs in a
 photon
fluctuation becomes
 eventually large at very low $x$.
 However, only one of
these virtual $\bar B$s has a
 chance to
 annihilate with the target proton
and
 create a baryon asymmetry with
 corresponding $x$.
 Thus, annihilation
of different $B\bar B$ pairs in
 the
 photon fluctuation shadow each
other.
 
 In order to take into account
 the growth of the
 integral in
eq.~(\ref{19}) we should introduce
 unitarity
 correction and renormalize the
BN flow
 
 \beq
 \widetilde\Delta_B(x) =
 \Delta_B(x)\left[\ 
\int\limits_{x_{min}}^1
 \frac{dx'}{x'}\Delta_B(x')
 \right]^{-1}
\label{21}
\eeq
 
 \noi
 This expression obviously satisfy the sum rule
eq.~(\ref{19}). It is easily interpreted: the
 total amount of the net BN,
which flows down to
 low $x$ is fixed, but phase space $\ln(1/x_{min})$
where this BN can be distributed grows with
 energy. Therefore, $\Delta_B(x)$
and $A_B(x)$ must
 decrease with energy at fixed $x$, what is
 provided
with
 the denominator in (\ref{21}).
 Nevertheless, due to the smallness of
the asymptotic
 value
 of BN flow estimated in (\ref{16}) the unitarity
correction
 (\ref{21}) is quite small at presently available
 energies. For the biggest rapidity interval of HERA
all the curves shown in Fig.~5 must be renormalized 
by about $20\%$ down due to the unitarity corrections
(\ref{21}).
\vspace{0.7cm}
 
\noindent
 {\large\bf 6. Discussion and conclusions.}
\vspace{0.3cm}
 
 A proton looks in its infinite-momentum frame like
 a
cloud of partons, quarks, antiquarks and gluons.
 The question, how the BN
of
 the proton is
 distributed in such a cloud is the main issue of
 the
present
 paper. Our main observations are:
 
\begin{itemize}

\item
 BN of the proton can be carried either by the
 valence quarks or by
 the sea quarks and gluons.
 In the latter case BN is distributed like $1/x$
 down to very low $x$.  We predict an unusual
 phenomenon, baryon asymmetry
 of
 the sea in the
 proton at very low $x$, which we estimate at $A_B(x)
 \approx
 0.07$. This number reflects the
 admixture of the
 $BN-\overline{BN}$
 exchange in
 the Pomeron. Unitarity corrections suppress
$A_B(x)$ dependent on the rapidity interval of
the $\gamma^*p$ collision. This is $20\%$ effect
for the energy range of HERA.
 
\item
 The BN distribution at medium $x$ is provided by a
 single valence
 quark. It is the dominant
 contribution to the baryon asymmetry $A_B(x)
 \approx
 3/\sqrt{x}$ down to $x \approx 5\times 10^{-4}$.
 
\item
 The baryon asymmetry at the parton level can be
 observed through a
 baryon asymmetry of produced
 particles in proton interactions at
 high-energies.
 The smallest $x\approx 10^{-5}$ can be reached in
 (virtual)
 photon - proton interactions at HERA.
 
\end{itemize}

 An important ingredient of our consideration is
 the method, which is
 used
 for the calculation of the
 BN distribution.  It is based on
 Lorenz-invariance
 of the observable baryon asymmetry, and allows one
 to
 replace the problem of
 BN distribution in the
 projectile proton by the
 rather well known process
 of annihilation of the primordial anti-BN with
 the
 target proton. In this way
 we derived
 eq.~(\ref{9b}), which is the
 central result of the
 paper. Using
 it, we predicted the baryon
 asymmetry
 provided by the valence quarks and
 gluons, which is shown for kinematics of
 HERA in
 Fig.~5.
 
 It is worth while
 reminding that the quark mechanism
 of BN transfer suggested in \cite{kz4}
 is
 different and provides much
 larger baryon
 asymmetry than what follows
 from the baryon
 asymmetry in
 hadronization of a valence quark.
 
 Although
 we use the ideas and the
 results of
 \cite{k}-\cite{kz6} and \cite{kz3,kz4}, 
essentially based on
 perturbative QCD
 calculations, which applicability
 is questionable, our
 predictions are
 free of this
 uncertainty.  In the case of the quark
 mechanism
 of BN
 transfer we use in (\ref{9b}) experimentally
 measured value
 of
 $\sigma^{p\bar p}_{ann}$. As for the
 gluonic mechanism, no direct measurement
 of baryon
 annihilation at high energies was done so far.
 However, we
 consider the asymptotic value of 
 $\sigma^{p\bar p}_{ann}\approx 1.5\ mb$ as
 a
 very
 reliable one, since it follows from the
 phenomenological analysis
 of
 data on multiparticle
 production \cite{kz2}, as well as from
 nonperturbative \cite{gn} and perturbative
 \cite{kz1} QCD estimations.
 
\medskip
 
 Experimental study of BN transfer through the
 biggest
rapidity
 interval was done so far at ISR
 \cite{isr1,isr2,isr-rev}.  It was
a
 measurement of
 the difference of inclusive cross sections of $p$
 and
$\bar
 p$ produced in central rapidity region.
 The smallest value of $x$
reached in
 this
 experiment was $x\approx 10^{-2}$.  
Unlike the proton colliders, one can use the 
whole rapidity interval with $e-p$ colliders. At HERA
it corresponds to 
$x\approx 10^{-5}$,
 which is smaller than at any
 of available or planned
proton colliders.
 
 We predict a flavour-independent baryon asymmetry.
 As
for the absolute
 production rate, we do not
 expect the usual suppression of
hyperon
 production
 compared with nucleons.  This is because the
 produced
baryons do
 not contain any light
 spectator quarks, but only string
junctions.  So,
 each of three quarks which joins the string
 junction to
build up the baryon
 may be either a
 strange quark or a light one. This
provides a
 combinatorial factor of three, which essentially
 compensates the
suppression
 for the strange quark
 production.  This is confirmed by nearly
the same
 branchings of $J/\Psi$ decay into $p\bar p$,
 $\Sigma \bar\Sigma$
and $\Xi
 \bar\Xi$
 \cite{tables}. This decay proceeds through three
gluons, which
 create a string
 junction-antijunction pair in final state,
which
 then
 dresses up with $u,\ d$ or $s$ quarks. Note
 that thanks to the
kinematics of
 HERA the baryons
 we are interested in, which are produced not
far
 from the
 photon fragmentation region, are not very
 energetic, what
makes the
 identification easier.
 One can study the production asymmetry
for
 $\Lambda$-hyperons, which may be easier
 identified.
 
\medskip
 
 {\bf Acknowledgements}: We would like to thank
 E.~Gabathuler,
T.~Greenshaw and D.~Milstead  for
informing us
 on preliminary results of the H1 Collaboration 
on $\Lambda \bar\Lambda$
 production in DIS
and H.~Meyer for
 pointing out the role of the baryon junction
in
 the decay of $J/\Psi$.  We are grateful to
 A.W.~Thomas for useful discussion and 
comments and to J.~Pochodzalla, who read the manuscript 
and made valuable improving suggestions.

\end{document}